\documentclass{article}[11pt]
\usepackage{amsfonts}
\usepackage{CJK,amsmath,indentfirst}
\usepackage{graphics}
\usepackage{graphicx}
\usepackage{amsmath}
\usepackage{amssymb}
\begin{document}

\title{\bf Primary branch solutions of first order autonomous scalar partial differential equations}

\author{\footnotesize S. Y. Lou$^{1,2}$\thanks{Email: lousenyue@nbu.edu.cn}  
 and R. X. Yao$^{3}$\\
\footnotesize $^{1}$ \it Shanghai Key Laboratory of Trustworthy Computing, East China Normal University, Shanghai 200062, China,\\
\footnotesize $^{2}$\it Ningbo Collabrative Innovation Center of Nonlinear Harzard System of Ocean and Atmosphere\\
\footnotesize \it and Faculty of Science, Ningbo University,  Ningbo, 315211, China\\
\footnotesize $^{3}$\it School of Computer Science, Shaanxi Normal University, Xi'an, 710062, China
}
\date{}
\maketitle
\parindent=0pt
\textbf{Abstract:} A primary branch solution (PBS) is defined as a solution with $n$ independent $m-1$ dimensional arbitrary functions for an $n$ order $m$ dimensional partial differential equation (PDE). PBSs of arbitrary first order scalar PDEs can be determined by using Lie symmetry group approach. 
Especially, one recursion operator and some sets of infinitely many high order symmetries are also explicitly given for arbitrary (1+1)-dimensional first order autonomous PDEs. 
Because of the intrusion of the arbitrary function, various implicit special exact solutions can be find by fixing the arbitrary functions and selecting different seed solutions. 
  \\ \\
\textbf{Mathematics Subject Classification:} 35F20, 35D99 
\\

\vskip.4in
\renewcommand{\thesection}{\arabic{section}}
\parindent=20pt

\section{Introduction}

It is well known that the symmetry group theory is very useful while studying exact solutions of nonlinear systems. In fact, group theory was first established by Abel and Galois to prove the nonexistence of general algebraic solution for a five degree univariate polynomial equation. Sophus Lie and Felix Klein introduced Lie symmetry algebras to study partial differential equations(PDEs)\cite{Olver}. In this paper, we try to find primary branch solutions (PBSs) of arbitrary autonomous first order single component partial differential equations in any dimensions. \\
\bf Definition 1. \rm \em For an $n$ order $m$ dimensional PDE, a solution is called a PBS if it contains $n$ independent $m-1$ dimensional arbitrary functions.
If only $n_1<n$ independent $m-1$ dimensional arbitrary functions are included in a solution, then it is called an secondary branch solution (SBS). If a solution contains some lower dimensional ($m_1<m-1$ dimensional) arbitrary functions, then it is called a degenerate solution. \rm

In fact, PBSs exist for all Painlev\'e integrable systems. The singularity analysis (Painlev\'e analysis) shows that if a system of PDEs is Painlev\'e integrable, then there exists a primary branch such that enough arbitrary functions (same as that in the definition of the PBSs) are included in a formal series solution with respect to an arbitrary singularity manifold \cite{PP1,PP2,PP3,PP4}. Thus, if one tries to find the PBSs of a PDE system by means of symmetry theory, then enough symmetries with enough arbitrary functions need to be found. Conversely, if we can find enough symmetries with enough arbitrary functions, then, we may successfully find the PBSs of the related PDE system. To make this idea clear, we study the symmetries and exact solutions of the following arbitrary ($n+1$)-dimensional first order PDE,
\begin{equation}\label{Fn}
F(u,\ u_0,\ u_{1},\ \ldots,\ u_{n})=0,\quad \ u_i\equiv \frac{\partial u}{\partial x_i},\quad i=0,\ \ldots,\ n,
\end{equation}
where $F$ is an arbitrary function and $n$ is an arbitrary positive integer. For simplicity, $u_i, (i=0,\ 1,\ ...,\ n)$ denote the partial differentiations with respect to $t(\equiv x_0),\ x_1(\equiv x),\ x_2,\ \ldots,\ x_n$ respectively.

Obviously, there exists only one $n$ dimensional arbitrary function for the PBSs of the first order PDE \eqref{Fn}. Thus, the SBSs of \eqref{Fn} are also degenerate solutions. In this paper we do not discuss degenerate solutions of \eqref{Fn}. Usually, degenerate solutions of \eqref{Fn} can be recasted to PBSs for some suitable lower dimensional models.

It is worth to emphasize that various special cases of \eqref{Fn} can be widely found in many physical fields. For instance, the Hopf ($b=0$) \cite{hopf} and damped Hopf ($b\neq0$) \cite{hpf} equations (also named Burgers and damped Burgers equations)
\begin{equation}\label{Hopf}
u_0=a uu_1+b u,
\end{equation}
with arbitrary constants $a$ and $b$
is a standard model to describe shock waves \cite{hpf1} with and without damping.

A general Hopf equation, a model equation of gas dynamics,
\begin{equation}\label{GHopf}
u_0=f(u)u_1,\qquad (f(u)\ \mbox{\rm being an arbitrary function of $u$}),
\end{equation}
is also widely used in hydrodynamics, multiphase flows, wave theory, acoustics, chemical engineering and other applications \cite{handb,para}.

The most general two dimensional form of \eqref{Fn} with the form
\begin{equation}\label{GHpf}
F(u,\ u_1,\ u_2)=0
\end{equation}
defines cylindrical surfaces whose elements are parallel to the $\{x_1x_2\}$ plane \cite{EK}.

A simple multiple dimensional significant model,
\begin{equation}\label{game}
\sum_{i=0}^n a_iu_i^2=c, \qquad c,\ a_i,\ (i=0,\ 1,\  \ldots, n),\ \mbox{being arbitrary constants},
\end{equation}
is encountered in differential games\cite{game}.

The paper is organized as follows. The symmetries and exact solutions of the general 1+1 dimensional autonomous systems are studied in Section 2 by introducing the idea of invariant functions. Additionally, a recursion operator for the arbitrary 1+1 dimensional first order autonomous system is also given in Section 2. Using the same idea as in Section 2, the symmetries and exact PBSs of the first order arbitrary autonomous system in any dimensions are investigated in Section 3. The last section is a short summary and discussion.

\section{Symmetries and solutions of (1+1)-dimensional systems}
Prior to study the general first order PDE \eqref{Fn}, we investigate the symmetries and exact solutions of Eq. \eqref{Fn} in (1+1)-dimensional case. Under this case,  Eq. \eqref{Fn} may have several non-degenerate branches which can be written in the following form
\begin{equation}\label{Bi}
u_0=B_i(u,\ u_1),\quad i=1,\ 2,\ \ldots,\ m.
\end{equation}
For every one branch given in \eqref{Bi}, we rewritten it as
\begin{equation}\label{Be}
u_0=F(u,\ u_1)u_1,
\end{equation}
where one $u_1$ factor has been separated from any given function $F(u,\ u_1)$ for simplicity later.

A symmetry, denoted by $\sigma$, of Eq. \eqref{Be} is defined as a solution of its linearized equation,
\begin{equation}\label{Bes}
\sigma_0-F_uu_1\sigma-(u_1F)_{u_1}\sigma_1=0,
\end{equation}
which means that Eq. \eqref{Be} is invariant under the transformation
\begin{equation}\label{trans}
u\rightarrow u+\epsilon \sigma
\end{equation}
with infinitesimal parameter $\epsilon$.

To look for the solution of the symmetry equation \eqref{Bes}, the following invariant function theorem will be very useful. \\
\bf Theorem 2.1. \rm \em If $\theta$ is a known symmetry of Eq. \eqref{Be}, and $\varphi$ is an invariant function which satisfies the invariant equation
\begin{equation}\label{Beg}
\varphi_0-(u_1F)_{u_1}\varphi_1=0,
\end{equation}
then
\begin{equation}\label{rs1}
\sigma=\varphi\theta
\end{equation}
is also a symmetry of Eq. \eqref{Be}. \rm \\
\it Proof. \rm Substituting Eq. \eqref{rs1} into
Eq. \eqref{Bes}, we have
\begin{eqnarray}\label{Bes1}
&&
(\varphi\theta)_0-u_1F_u\varphi\theta-(u_1F)_{u_1}(\varphi\theta)_1 \nonumber\\
&&=\varphi_0\theta+\varphi\theta_0-u_1F_u\varphi\theta
-(u_1F)_{u_1}(\varphi_1\theta+\varphi\theta_1)\nonumber\\
&&=\varphi[\theta_0-u_1F_u\theta-(u_1F)_{u_1}\theta_1]
+\theta[\varphi_0
-(u_1F)_{u_1}\varphi_1]=0,
\end{eqnarray}
Due to the fact that $\theta$ is a symmetry and $\varphi$ is an invariant function, thus, Theorem 2.1 is proved. $\blacksquare$

Following Theorem 2.1, the problem of finding symmetries of Eq. (2) is converted into that of finding its invariant functions. To this end, the Theorem below will significantly simplify the procedure.\\
\bf Theorem 2.2. \rm \em If $\varphi'$ is an invariant function, so is
\begin{equation}\label{rv}
\varphi=g(\varphi'),
\end{equation}
where $g\equiv g(\varphi')$ is some arbitrary function of $\varphi'$.\rm \\
\it Proof. \rm  Substituting Eq. \eqref{rv} into Eq. \eqref{Beg}, we have
\begin{eqnarray}\label{Beg1}
&&g_0-(u_1F)_{u_1}g_1\nonumber\\
&&=g_{\varphi'}[\varphi'_0-(u_1F)_{u_1}\varphi'_1]
\nonumber\\
&&=0.
\end{eqnarray}
Because $\varphi'$ is a group invariant function, the last step of \eqref{Beg1} holds. Thus, Theorem 2.2 is proved. $\blacksquare$

Based on the Theorems 2.1 and 2.2, if we can find some types of nontrivial (nonconstant) group invariant functions and some fixed symmetries, then new types of symmetries with arbitrary functions can be constructed.

To find more symmetries for a nonlinear PDE, we use the following definitions on the strong symmetry operator (SSO) and the recursion operator (RO).\\
\bf Definition 2. \rm \em An operator $\Phi$ is called an SSO of a given PDE if it transforms a symmetry $\sigma'$ to a new symmetry $\Phi\sigma'$. An SSO is called RO if it is also a hereditary operator (HO) which satisfies \cite{HO}
\begin{eqnarray}\label{Q}
\Phi'[\Phi f] g - \Phi'[\Phi g] f   -\Phi (\Phi'[f]g-\Phi'[g]f )=0,
\end{eqnarray}
 for arbitrary $f$ and $g$.
Here the Fr\'echet derivative of $\Phi$ is defined by \[\Phi'[h]\equiv \frac{d}{d\varepsilon}\Phi(u+\varepsilon h)\mid_{\varepsilon=0},\]
for any $h$.\rm

In (1+1)-dimensional case, it is known that if a nonlinear system is integrable, one can find one or more recursion operators. Since looking for a symmetry \eqref{Be} is equivalent to finding its invariant function, we define the invariant operator $\phi$  which transforms an invariant function $\varphi'$ to a new one $\phi\varphi'$.

For Eq. \eqref{Be}, we have the following theorem to provide an invariant operator.\\
\bf Theorem 2.3. \rm \em If $\varphi'$ is an invariant function, so is
\begin{equation}\label{rv1}
\varphi=\phi\varphi'
\end{equation}
with invariant operator
\begin{equation}\label{rphi}
\phi=\frac{1}{G_uu_1+G_{u_1}u_{11}}\partial_x=\frac1{G_1}\partial_x,
\end{equation}
where $G\equiv G(u,\ u_1)$ is related to $F$ by
\begin{equation}\label{K12}
F_{u_1}G_{u}-F_uG_{u_1}=cu_1^{-2},
\end{equation}
with $c$ being an arbitrary constant.
 \rm\\
\it Proof. \rm 
In terms of \eqref{K12}, we have 
\begin{eqnarray}
G_0&=&G_1(u_1F)_{u_1}-c,\nonumber\\
G_{10}&=&G_{11}(u_1F)_{u_1}+G_1\big[(u_1F)_{u_1}\big]_1.\label{G10}
\end{eqnarray}
Substituting Eqs. \eqref{rv1}-\eqref{rphi} with \eqref{G10} into Eq. \eqref{Beg}, we have
\begin{eqnarray}\label{Bega}
&&\left(\frac{\varphi'_1}{G_1}\right)_0
-(u_1F)_{u_1}\left(\frac{\varphi'_1}{G_1}\right)_1\nonumber\\
&&=\frac{\varphi'_{10}}{G_1}-\frac{G_{10}}{G_1^2}\varphi'_1
 -(u_1F)_{u_1}\left(
\frac{\varphi'_{11}}{G_1}-\frac{G_{11}}{G_1^2}\varphi'_1
\right)\nonumber\\
&&=\frac{(u_1F)_{u_1}\varphi'_{11}+\big[(u_1F)_{u_1}\big]_1\varphi'_1}{G_1}-\frac{G_{11}(u_1F)_{u_1}+G_1\big[(u_1F)_{u_1}\big]_1}{G_1^2}\varphi'_1\nonumber\\
&&\quad-(u_1F)_{u_1} \left(
\frac{\varphi'_{11}}{G_1}-\frac{G_{11}}{G_1^2}\varphi'_1
\right)\nonumber\\
&&=0.
\end{eqnarray}
Theorem 2.3 is proved. $\blacksquare$

Because \eqref{K12} is only a 1+1 dimensional linear first order PDE for any given $F$, it is straightforward to find its general solution in the form 
\[G=\left\{\begin{array}{ll}A(F)-\int^u \frac{c}{B^2 F_B(y,B)}dy, &  F_{u_x}\neq 0,\\
A(F)+\frac{c}{u_x F_u}, &  F_{u_x}=0,\ F_u\neq 0,\\
\mbox{arbitrary function of $\{u,\ u_x\}$}, & F_u=F_{u_x}=c=0,
\end{array}\right.\]
where  $B\equiv B(y,\ u,\ u_x)$ is determined by $F(y,B)=F(u,u_x )$ and $A$ is an arbitrary function of $F$.

Using the Theorems 2.1 and 2.3, one of the RO of \eqref{Be} related to the invariant operator \eqref{rphi} can be found immediately.\\
\bf Theorem 2.4. \rm \em If $\sigma'$ is a symmetry of  Eq. \eqref{Be}, so is
\begin{eqnarray}\label{Rs}
\sigma=\Phi\sigma'
\end{eqnarray}
with RO
\begin{eqnarray}\label{RO}
\Phi =u_1\phi \frac1{u_1}=\frac{u_1}{G_1}\partial_x u_1^{-1}.
\end{eqnarray}\rm
\it Proof. \rm One can directly prove $\Phi$ given in \eqref{RO} is an SSO of \eqref{Be} by substituting Eq. \eqref{Rs} into the symmetry definition equation \eqref{Bes}. However, it is much simpler to rewrite the invariant functions of the theorem 2.3, $\varphi'$ and $\varphi$, as
\begin{eqnarray} \label{vs}
\varphi' =\sigma'/u_1,\quad \varphi =\sigma/u_1,
\end{eqnarray}
because of Theorem 2.1 and $u_1$ being a trivial symmetry related to $x$ translation.
Substituting Eqs. \eqref{vs} into Theorem 2.3 leads to the conclusion that $\Phi$ given by \eqref{RO} is an SSO of \eqref{Be}. To prove the theorem, we need further to prove $\Phi$ is also an HO that means we should check \eqref{Q}
is identically satisfied with arbitrary $f$ and $g$ for $\Phi$ given by 
\eqref{RO}. 

Substituting \eqref{RO} into \eqref{Q}, we get the following formula from the left hand side of \eqref{Q}  
\begin{eqnarray}
&&\left[\frac{(\Phi f)_1}{G_1}-\frac{u_1(G'_1\Phi f)}{G_1^2}\right]\left(\frac{g}{u_1}\right)_1-\frac{u_1}{G_1}\left[\frac{g(\Phi f)_1}{u_1^2}\right]_1-\frac{u_1}{G_1}\left[\frac{f_1G_1-u_1(G'_1f)}{u_1G_1^2}\left(\frac{g}{u_1}\right)_1-\frac{1}{G_1}\left(\frac{gf_1}{u_1^2}\right)_1\right]_1\nonumber\\
&&-\left[\frac{(\Phi g)_1}{G_1}-\frac{u_1(G'_1\Phi g)}{G_1^2}\right]\left(\frac{f}{u_1}\right)_1+\frac{u_1}{G_1}\left[\frac{f(\Phi g)_1}{u_1^2}\right]_1+\frac{u_1}{G_1}\left[\frac{g_1G_1-u_1(G'_1g)}{u_1G_1^2}\left(\frac{f}{u_1}\right)_1-\frac{1}{G_1}\left(\frac{fg_1}{u_1^2}\right)_1\right]_1.\nonumber\\ \label{Q2}
\end{eqnarray}
Because of 
\begin{equation}\label{rG1}
G_1=G_uu_1+G_{u_1}u_{11}, \quad \Phi f=\frac{u_1}{G_1}\left(\frac{f}{u_1}\right)_1,
\end{equation}
we have 
\begin{eqnarray}
G_{11}&=&G_{u_1u_1}u_{11}^2+G_{uu}u_1^2+2G_{uu_1}u_1u_{11}+G_{u}u_{11}+G_{u1}u_{111},\nonumber\\
G_{111}&=&G_uu_{111}+G_{u_1}u_{1111}+3G_{uu}u_1u_{11}+3(u_1u_{111}+u_{11}^2)G_{uu_1}+3u_{11}(u_{111}G_{u_1u_1}+u_1^2G_{uuu_1}\nonumber\\
&&+u_1u_{11}G_{uu_1u_1})+u_1^3G_{uuu}+G_{u_1u_1u_1}u_{11}^3,\nonumber\\
G'_1f&=&(G_{uu} u_1+G_{uu_1}u_{11})f+(G_u+u_1G_{uu_1}+u_{11}G_{u_1u_1})f_1+G_{u_1}f_{11},\nonumber\\
G'_1\Phi f&=&(G_{uu} u_1+G_{uu_1}u_{11})\frac{u_1}{G_1}\left(\frac{f}{u_1}\right)_1+(G_u+u_1G_{uu_1}+u_{11}G_{u_1u_1})\left[\frac{u_1}{G_1}\left(\frac{f}{u_1}\right)_1\right]_1\nonumber\\
&&+G_{u_1}\left[\frac{u_1}{G_1}\left(\frac{f}{u_1}\right)_1\right]_{11}.\label{G11}
\end{eqnarray}
Substituting the relations \eqref{rG1}, \eqref{G11} and the similar relations with the exchanges $f\leftrightarrow g$ into \eqref{Q2} and expanding all the differentiations one can find that \eqref{Q2} is exactly zero for arbitrary $f$ and $g$. Thus the SSO $\Phi$ is also an HO, which means that it is really an RO of \eqref{Be}. 
The theorem 2.4 is proved. $\blacksquare$
\\
\bf Remark. \rm Because of the hereditary property of $\Phi$ given by Eq. \eqref{RO}, we get a general integrable hierachy 
\begin{equation}\label{un}
u_t=\Phi^n (u_1F),\ 
\end{equation}
with infinitly many commute symmetries $K_m\equiv \Phi^m (u_1F),\ m=1,\ 2,\ \ldots.$ We do 
not discuss the properties of the hierary  \eqref{un} in this paper.

Now it is ready to write down the following general symmetry theorem.\\
\bf Theorem 2.5. \rm \em The (1+1)-dimensional autonomous  equation \eqref{Be} possesses the generalized symmetries
\begin{eqnarray}\label{gs}
\sigma=S(\alpha_n,\ \beta_m,\ \gamma_k,\ n,m,k=0,\ 1,\ \ldots)u_0
\end{eqnarray}
with
\begin{subequations}
\begin{eqnarray}\label{gsa}
\alpha_n&\equiv& \phi^n \frac{u_1}{u_0}, \\
\label{gsb}
 \beta_m&\equiv& \phi^m (ax+A),  \quad u_1^2(A_{u_1}F_u-A_uF_{u_1})=a(u_1F)_{u_1},
 \\
\label{gsc}
 \gamma_k&\equiv& \phi^k(bt+B), \quad u_1^2(B_{u_1}F_u-B_uF_{u_1})=-b,
\end{eqnarray}
\end{subequations}
and $S$ being an arbitrary function of the arguments $\alpha_n,\ \beta_m$ and $\gamma_k$.
 \rm \\
\bf Proof. \rm Because of the Theorems 2.1, 2.2 and 2.3, the only thing we need to do is to prove $\beta_0\equiv ax+A(u,\ u_1)$ and $\gamma_0\equiv t+B(u,\ u_1)$ are invariant functions of \eqref{Be} while $\alpha_0$ is trivially invariant function owing to $u_0$ and $u_1$ being two trivial symmetries related to time and space translation invariance. Thus, after substituting $\beta_0$ into the invariant function equation \eqref{Beg}, we have
\begin{eqnarray}\label{gA}
0&=&(x+A)_0-(u_1F)_{u_1}(x+A)_1\nonumber\\
&=&A_uu_0+A_{u_1}u_{10}-(u_1F)_{u_1}(a+u_1 A_u+u_{11}A_{u_1})\nonumber\\
&=&A_uFu_1+A_{u_1}(u_1^2F_u+u_{11}(u_1F)_{u_1})-(u_1F)_{u_1}(a+u_1 A_u+u_{11}A_{u_1})\nonumber\\
&=&u_1A_uF+u_1[A_{u_1}u_1F_u-A_u(F+u_1F_{u_1})]-a(u_1F)_{u_1})
\nonumber\\
&=&u_1^2[A_{u_1}F_u-
A_uF_{u_1}]-a(u_1F)_{u_1}.
\end{eqnarray}
The last step is just the definition of $A$, therefore, $\beta_0=ax+A$ is proved to be an invariant function. In the same way, one can prove that $\gamma_0=bt+B$ is an invariant function.  Theorem 2.5 is proved. $\blacksquare$

The definition equations of $A$ and $B$ 
are only the first order linear equations of $A$ and $B$ with respect to the independent variables $u$ and $u_1$. The first order linear equation can be solved by means of the standard characteristic method. The general solutions of Eqs. \eqref{gsb}-\eqref{gsc} take the forms
\begin{eqnarray}
\label{rA}
A&=&\left\{ \begin{array}{ll}
-a\displaystyle{\int^u \frac{F+yF_y(b,y)}{y^2 F_y(b,y)}\mbox{\rm d}b+A_0(F),} & F_{u_1}\neq0, \cr
-a\displaystyle{\frac{F}{u_1F_u}+A_0(F),} & F_{u_1}=0,\ F_u\neq0, \cr
A_0(u,\ u_1), & F_{u_1}=F_u=a=0,
\end{array}\right.
 \\
 \label{rB}
B&=&\left\{ \begin{array}{ll}
b\displaystyle{\int^u \frac{1}{y^2 F_y(b,y)}\mbox{\rm d}b+B_0(F),} & F_{u_1}\neq0, \cr
b\displaystyle{\frac{1}{u_1F_u}+B_0(F),} & F_{u_1}=0,\ F_u\neq0, \cr
B_0(u,\ u_1), & F_{u_1}=F_u=b=0,
\end{array}\right.
\end{eqnarray}
where $A_0$ and $B_0$ are arbitrary functions of the indicated variables which can be simply taken as zero without loss of generality.

 As pointed out in Section 1, if a (1+1)-dimensional first order PDE can be solved by quadrature, a one-dimensional arbitrary function must be included in its PBSs. From the Theorem 2.5, an arbitrary function with arbitrary arguments have been included, which implies that the symmetries have been over-determined to find its PBSs. In fact, to find some PBSs of the (1+1)-dimensional first order PDE \eqref{Be}, it is enough to use a symmetry with a \rm one\rm-dimensional arbitrary function, say,
\begin{equation}\label{a0}
\sigma=g(\alpha_0)u_0,\ \alpha_0\neq\mbox{\rm constant},
\end{equation}
or,
\begin{equation}\label{b0}
\sigma=h(\beta_0)u_0,\ \alpha_0=\mbox{\rm constant},\ \beta_0\neq\mbox{\rm constant}.
\end{equation}

Using the symmetry expression \eqref{a0}, we have the following solution theorem.\\
\bf Theorem 2.6. \rm\em If $u=U(t,\ x)$ is a seed solution of \eqref{Be} with the conditions $U_xU_t\neq 0$ and $U_x/U_t\neq {\rm constant}$, then its PBS has the form
\begin{equation}\label{ru}
u'(t,x)=U(t'(t,x), x'(t,x))
\end{equation}
with $t'\equiv t'(t,\ x),\ x'\equiv x'(t,\ x)$ being determined by
\begin{subequations}\label{xt}
\begin{eqnarray}
&&t'=t-g(\eta)+\eta g_{\eta},\\
&&x'=x- g_{\eta},\
\\
&&\eta\equiv \left.\frac{U_1}{U_0}\right|_{t\rightarrow t',\ x\rightarrow x'}.
\end{eqnarray}
\end{subequations}
\it Proof. \rm Firstly, we calculate the Jacobian determinant of the transformation from $\{x,\ t\}$ to $\{x',\ t'\}$
\begin{eqnarray}\label{JB}
\Delta&\equiv &\frac{J(x',\ t')}{J(x,\ t)}\nonumber\\
&=&x'_1t'_0-x'_0t'_1.
\end{eqnarray}
From \eqref{xt}, we have
\begin{eqnarray}\label{xt10}
&&t'_1=-g_{\eta}\eta_1+\eta_1g_{\eta}+\eta g_{\eta\eta}\eta_1,\nonumber\\
&&x'_1=1- g_{\eta\eta}\eta_1,\nonumber\
\\
&&t'_0=1-g_{\eta}\eta_0+\eta_0g_{\eta}+\eta g_{\eta\eta}\eta_0,\nonumber\\
&&x'_0=- g_{\eta\eta}\eta_0,
\end{eqnarray}
and
\begin{eqnarray}\label{eta10}
\eta_1&=&\left(\frac{U_1}{U_0}\right)_1\nonumber\\
&=&\frac{U_{11}x'_1+U_{10}t'_1}{U_0}
-\frac{U_1(U_{10}x'_1+U_{00}t'_1)}{U_0^2},\nonumber\\
\eta_0&=&\left(\frac{U_1}{U_0}\right)_0\nonumber\\
&=&\frac{U_{11}x'_0+U_{10}t'_0}{U_0}
-\frac{U_1(U_{10}x'_0+U_{00}t'_0)}{U_0^2}.
\end{eqnarray}
Substituting \eqref{eta10} into \eqref{xt10}, we have,
\begin{eqnarray}\label{xt11}
&&t'_1=\eta g_{\eta\eta}\left(\frac{U_{11}x'_1+U_{10}t'_1}{U_0}
-\frac{U_1(U_{10}x'_1+U_{00}t'_1)}{U_0^2}\right),\nonumber\\
&&x'_1=1- g_{\eta\eta}\left(\frac{U_{11}x'_1+U_{10}t'_1}{U_0}
-\frac{U_1(U_{10}x'_1+U_{00}t'_1)}{U_0^2}\right),\nonumber\
\\
&&t'_0=1+\eta g_{\eta\eta}\left(\frac{U_{11}x'_0+U_{10}t'_0}{U_0}
-\frac{U_1(U_{10}x'_0+U_{00}t'_0)}{U_0^2}\right),\nonumber\\
&&x'_0=- g_{\eta\eta}\left(\frac{U_{11}x'_0+U_{10}t'_0}{U_0}
-\frac{U_1(U_{10}x'_0+U_{00}t'_0)}{U_0^2}\right),
\end{eqnarray}
which on solving about $\{x'_1,\ x'_0,\ t'_1,\ t'_0\}$ arrives at
\begin{eqnarray}\label{rxt}
x'_1&=&\frac{U_0^3-\delta_1}{\delta},\quad \delta_1\equiv(U_0U_{10}-U_1U_{00})U_1g_{\eta\eta},\
\delta\equiv U_0^3+(U_0^2U_{11}+U_{00}U_1^2-2U_0U_1U_{10})g_{\eta\eta},  \nonumber\\
t'_1&=&\frac{\delta_2}{\delta},\quad \delta_2\equiv(U_0U_{11}-U_1U_{10})U_1g_{\eta\eta},\ \nonumber\\
x'_0&=&-\frac{U_0\delta_1}{U_1\delta}, \nonumber\\
t'_0&=&\frac{U_0(\delta_2+U_0^2U_1)}{U_1\delta}.
\end{eqnarray}
Substituting \eqref{rxt} into \eqref{JB}, we have
\begin{eqnarray}\label{Jb}
\Delta&=&\frac{U_0^3}{\delta}\neq 0,
\end{eqnarray}
which means that the transformation \eqref{xt} is a one to one mapping.

Now we calculate the differentiations of the new solution $u'(x,\ t)$,
\begin{eqnarray}\label{u'xt}
u'_1&=&\partial_x U(x'(x,t),t'(x,t))\nonumber\\
&=&U_1x'_1+U_0t'_1\nonumber\\
&=&U_1\frac{U_0^3-\delta_1}{\delta}
+U_0\frac{\delta_2}{\delta}\nonumber\\
&=&U_1\frac{U_0^3-(U_0U_{10}-U_1U_{00})U_1g_{\eta\eta}}
{U_0^3+(U_0^2U_{11}+U_{00}U_1^2-2U_0U_1U_{10})g_{\eta\eta}}
+U_0\frac{(U_0U_{11}-U_1U_{10})U_1g_{\eta\eta}}{U_0^3+(U_0^2U_{11}
+U_{00}U_1^2-2U_0U_1U_{10})g_{\eta\eta}}\nonumber\\
&=&U_1,\\
u'_0&=&\partial_t U(x'(x,t),t'(x,t))\nonumber\\
&=&U_1x'_0+U_0t'_0\nonumber\\
&=&U_1\frac{-\delta_1U_0}{\delta U_1}
+U_0\frac{U_0(\delta_2+U_0^2U_1)}{U_1\delta}\nonumber\\
&=&-U_0\frac{(U_0U_{10}-U_1U_{00})U_1g_{\eta\eta}}
{U_0^3+(U_0^2U_{11}+U_{00}U_1^2-2U_0U_1U_{10})g_{\eta\eta}}
+U_0^2\frac{[(U_0U_{11}-U_1U_{10})g_{\eta\eta}+U_0^2]}
{U_0^3+(U_0^2U_{11}
+U_{00}U_1^2-2U_0U_1U_{10})g_{\eta\eta}}\nonumber\\
&=&U_0.
\end{eqnarray}
Thus,
\begin{eqnarray}\label{2'}
&&u'_0-F(u',\ u'_1)\nonumber\\
&&=U_0-F(U,\ U_1)\nonumber\\
&&=0,
\end{eqnarray}
Theorem 2.6 is proved. $\blacksquare$

Alternatively, Theorem 2.6 can be proved directly by solving the initial value problem related to the symmetry \eqref{a0} because the symmetry \eqref{a0} can be rewritten as
\begin{equation}\label{a0'}
\sigma=\left(g-\alpha g_{\alpha}
\right)u_0+g_{\alpha}u_1\equiv Tu_0+Xu_1,
\end{equation}
while the corresponding first order prolongation ${\rm pr}^{(1)}\sigma$ is a closed one,
\begin{equation}
{\rm pr}^{(1)}\sigma=\left(g-\alpha g_{\alpha}
\right)\partial_t+g_{\alpha}\partial_x+0\cdot \partial_u
+0\cdot \partial_{u_0}+0\cdot \partial_{u_1},
\end{equation}
which means that the symmetry \eqref{a0} (or equivalently \eqref{a0'}) is only a Lie point symmetry on the prolonged space $\{x,\ t,\ u,\ u_0,\ u_1\}$. Thus, based on the Lie's first principle, the finite transformation of the symmetry is determined by the following initial value problem,
\begin{subequations}\label{inv}
\begin{eqnarray}
&&\frac{\mbox{\rm d}t(\epsilon)}{\mbox{\rm d}\epsilon}=\left(g-\alpha g_{\alpha}\right)(\epsilon),\quad t(0)=t, \\
&&\frac{\mbox{\rm d}t(\epsilon)}{\mbox{\rm d}\epsilon}=g_{\alpha}(\epsilon),\quad x(0)=x,\\
&&\frac{\mbox{\rm d}u(\epsilon)}{\mbox{\rm d}\epsilon}=0,\quad u(0)=U,\\
&&\frac{\mbox{\rm d}u_1(\epsilon)}{\mbox{\rm d}\epsilon}=0,\quad u_1(0)=U_1,\\
&&\frac{\mbox{\rm d}u_0(\epsilon)}{\mbox{\rm d}\epsilon}=0,\quad u_0(0)=U_0.
\end{eqnarray}
\end{subequations}
The solution of the initial value problem of \eqref{inv} is nothing but the one given in Theorem 2.6 after using some new notations. We omit the details here because we will illustrate the problem by a special case in Section 3.
\\
\bf Remark. \rm (i). For a given PDE (given $F$ in this paper), there may have several primary branches and secondary branches. For every given branch, one may use the symmetry theory (or other approaches) to find its PBS. (ii). For every given branch, using different seed solutions and different symmetries one may find PBSs in quite different forms. These formally different PBSs should be equivalent, however, it will be very difficult to prove the equivalence. (iii). The symmetry \eqref{b0} can also been used to find the PBS in a different form but we do not disscuss it further instead of giving a special example for fixed $F$. 
\\
\bf Example. \rm To end up this section, we offer a special example
\begin{equation}\label{ex}
u_0=uu_1^2.
\end{equation}

For the special model \eqref{ex}, one of the recursion operators has the form
\begin{equation}\label{P1}
\Phi_1=u_1 \phi_1 u_1^{-1}
\end{equation}
with $\phi_1$ being an invariant operator
\begin{equation}\label{p1}
\phi_1=u^{-3}u_{11}^{-1}\partial_x.
\end{equation}
Some special higher order symmetries can be expressed by \begin{equation}
\sigma=S\left(\phi_1^i \frac{u_1}{u_0},\ \phi_1^j (x-uu_1^{-1}),\ \phi_1^k (2t+u_1^{-2}),\ i=1,\ 2,\ \ldots,\ n_1,\ j=1,\ 2,\ \ldots,\ n_2,\ k=1,\ 2,\ \ldots,\ n_3\right)u_0,
\end{equation}
with $S$ being an arbitrary function of the indicated variables.

The model \eqref{ex} possesses a special non-traveling wave solution
\begin{equation}\label{u_0}
U=\frac{x}{\sqrt{-2t}}
\end{equation}
which can be directly verified or derived from a high order (second order) symmetry constraint.

By using Theorem 2.6 and the special solution \eqref{u_0}, a PBS of the toy model \eqref{ex} can be written down in an implicit form
\begin{equation}\label{gs1}
u=\frac{x'}{\sqrt{-2t'}},
\end{equation}
where $x'=x'(x,t)$ and $t'=t'(x,t)$ are determined by
\begin{subequations}\label{gS}
\begin{equation}\label{gS1}
x'=x-\varepsilon G_\eta, \qquad \eta\equiv -\frac{2{t'}}{{x'}},
\end{equation}
\begin{equation}\label{gs2}
t'=t-\varepsilon \left(G+\frac{2{t'}}{{x'}}G_\eta\right),
\end{equation}
\end{subequations}
with $G\equiv G(\eta)$ being an arbitrary function of $\eta$.

Usually, the PBSs given by Eq. \eqref{gs1} are implicit. Only for some simple selection of $G(\eta)$, \eqref{gs1} can be writen in explicit form, say, if we take
\begin{equation}\label{ge}
G(\eta)=\eta^2,
\end{equation}
then the solution of \eqref{gS} reads
\begin{equation}
x'=x+\frac{4 t}{x},\ t'=\frac{t}{x^2}(4t+x^2),
\end{equation}
and then the related special solution has the form
\begin{equation}
u=\sqrt{-2-\frac{x^2}{2t}}.
\end{equation}

The forms of the PBSs are closely dependent on the selections of seed solutions though they may be equivalent. 
 
For the special toy model \eqref{ex}, a quite special trivial travelling wave solution has the form,
\begin{equation}\label{seedtrv}
u=\sqrt{2(x+t)}. 
\end{equation}
For the travelling wave seed solution the symmetry \eqref{a0} can not be used to find any new solutions. However, other symmetries, say,
 the second type of symmetry \eqref{b0}, i.e.,
\begin{equation}\label{b01}
\sigma=S(x-uu_1^{-1})u_1, \quad x-uu_1^{-1}\neq \mbox{constant,}
\end{equation}
with $S\equiv S(x-uu_1^{-1})$ being an arbitrary function of $x-uu_1^{-1}$ can be used to find the following related PBS with the help of the seed \eqref{seedtrv}. 

In terms of the seed \eqref{seedtrv} and the symmetry \eqref{b01}, the related PBS can be written as 
\begin{equation}\label{sol2}
u=\frac{S(Y^{-1}(a+Y(-\xi-2t)))}{S(-\xi-2t)}\sqrt{2\xi+2t},
\end{equation}
where $Y(y)$ is an arbitrary function of $y$ and it is related to the symmetry function $S(y)$ by 
$$S(y)=\frac{1}{Y_y},$$
$Y^{-1}$ is the inverse function of $Y$, and $\xi \equiv \xi(x,t)$ is related to the space time $\{x,t\}$ implicitly by 
\begin{equation}\label{sol2x}
x=2(\xi+t)\frac{S(Y^{-1}(a+Y(-\xi-2t)))}{S(-\xi-2t)}+Y^{-1}(a+Y(-\xi-2t)). 
\end{equation}
More specifically, 
if we take $Y(y)=\sin(y)$, then we have a special solution 
\begin{equation}\label{sol20}
u=\frac{\cos(\xi+2t)}{\sqrt{1-(a-\sin(\xi+2t))^2}}\sqrt{2\xi+2t}
\end{equation}
with
\begin{equation}\label{sol2x0}
x=2(\xi+t)\frac{\cos(\xi+2t)}{\sqrt{1-(a-\sin(\xi+2t))^2}}+\arcsin(a-\sin(\xi+2t)). 
\end{equation}

\section{Symmetries and solutions of arbitrary (n+1)-dimensional first order autonomous PDEs}

Based on the idea mentioned in Section 1,  finding a PBS of \eqref{Fn} is equivalent to finding a symmetry including one (one is enough for a first order PDE) $n$-dimensional arbitrary function, say,
\begin{equation}
\sigma=\sigma(t,\ x_1,\ \ldots,\ x_n,\ u,\ G(\tau_1,\ \ldots,\ \tau_n) ),\label{sigma}
\end{equation}
where $G$ should be an arbitrary function of $n$ \em independent \rm variables $\tau_i=\tau_i(t,\ x_1,\ \ldots,\ x_n)$.

A symmetry of Eq. \eqref{Fn}, $\sigma$, is defined as a solution of its linearized equation
\begin{equation}
F'\sigma\equiv \left(F_u+\sum_{i=0}^nF_{u_i} \partial_{x_i}\right)\sigma \equiv F_u\sigma+\sum_{i=0}^nF_{u_i} \sigma_i=0.\label{Eq.Sym}
\end{equation}
As in the (1+1)-dimensional case, we have the following invariant function theorem. \\
\bf Theorem 3.1. \rm \em If $\varphi$ is an invariant function defined as a solution of
\begin{equation}\label{3phi}
\sum_{i=0}^nF_{u_i} \varphi_i=0,
\end{equation}
$\theta$ is a symmetry of Eq. \eqref{Fn}, then
\begin{equation}
\label{3phi1}
\sigma=\varphi\theta
\end{equation}
is also a symmetry of Eq. \eqref{Fn}.\rm \\
\it Proof. \rm Substituting Eq. \eqref{3phi1} into Eq. \eqref{Eq.Sym}, we have
\begin{eqnarray}\label{3phi3} &&F_u\varphi\theta+\sum_{i=0}^nF_{u_i} (\varphi\theta)_i\nonumber\\
&&=F_u\varphi\theta+\sum_{i=0}^nF_{u_i} (\varphi_i\theta+\varphi\theta_i)\nonumber\\
&&=\varphi\left(F_u\theta
+\sum_{i=0}^nF_{u_i}\theta_i\right)+\theta \sum_{i=0}^nF_{u_i} \varphi_i\nonumber\\
&&=0.
\end{eqnarray}
The last step of \eqref{3phi3} is true because of $\theta$ and $\varphi$ being a symmetry and an invariant function of the model \eqref{Fn} respectively. Theorem 3.1 is proved. $\blacksquare$

The following corollary gives an alternative statement of Theorem 3.1. \\
\bf Corollary 3.1. \rm \em If $\sigma$ and $\theta$ are symmetries of Eq. \eqref{Fn}, then
$$\varphi=\frac{\sigma}{\theta}$$
is an invariant function of Eq. \eqref{Fn} which is defined as a solution of Eq. \eqref{3phi}. \rm

To get more general symmetry of \eqref{Fn}, we have the following theorem. \\
\bf Theorem 3.2. \rm \em If $\varphi'$ is an invariant function of Eq. \eqref{Fn}, so is
\begin{eqnarray}\label{3G}
\varphi=G(\varphi')
\end{eqnarray}
with $G\equiv G(\varphi')$ being an arbitrary function of $\varphi'$. \rm\\
\it Proof. \rm Substituting Eq. \eqref{3G} into Eq. \eqref{3phi}, we have
\begin{eqnarray}\label{3phi1}
&&\sum_{i=0}^nF_{u_i} G(\varphi')_i=G_{\varphi'}\sum_{i=0}^nF_{u_i} \varphi_i=0.
\end{eqnarray}
Theorem 3.2 is proved. $\blacksquare$

Now, it is straightforward to prove the symmetry theorem for the general (n+1)-dimensional autonomous first order PDE \eqref{Fn}. \\
\bf Theorem 3.3. \rm \em Arbitrary autonomous first order PDE \eqref{Fn} possesses the following symmetry
\begin{eqnarray}\label{3s}
\sigma=G(\tau_1,\ \tau_2,\ \ldots,\ \tau_n,\ \varphi_0,\ \varphi_1,\ \varphi_2,\ \ldots,\ \varphi_n)u_0
\end{eqnarray}
with $G\equiv G(\tau_1,\ \tau_2,\ \ldots,\ \tau_n,\ \varphi_0,\ \varphi_1,\ \varphi_2,\ \ldots,\ \varphi_n)$ being an arbitrary functions of the indicated variables,
\begin{eqnarray}\label{3ss}
&&\tau_\alpha\equiv \frac{u_\alpha}{u_0},\quad \alpha=1,\ 2,\ \ldots,\ n,\nonumber\\
&&\varphi_i\equiv x_i+\int^{u_0}\frac{F_{u_i}(f,\ b,\ b\tau_1,\ b\tau_2,\ \ldots,\ b\tau_n)}{b F_u(f,\ b,\ b\tau_1,\ b\tau_2,\ \ldots,\ b\tau_n)}\mbox{\rm d}b\equiv x_i+A_i(u,\ u_0,\ u_1,\ \ldots,\ u_n),
\end{eqnarray}
where $f\equiv f(u,\ u_0,\ u_1,\ \ldots,\ u_n,\ b)$ is a solution of
\begin{eqnarray}\label{3f}
F(f,\ b,\ b\tau_1,\ \ldots,\ b\tau_n)=F(u,\ u_0,\ u_1,\ \ldots,\ u_n).
\end{eqnarray}\rm
\it Proof. \rm Because $u_i,\ i=0,\ 1,\ \ldots,\ n$ are symmetries of Eq. \eqref{Fn} related to the space-time translations, the only thing we have to do is to prove $\varphi_i$ are invariant functions of Eq. \eqref{Fn}. Substituting $\varphi=\varphi_i=x_i+A_i(u,\ u_0,\ u_1,\ \ldots,\ u_n)\equiv x_i+A_i$ into Eq. \eqref{3phi}, we have
\begin{eqnarray}\label{3phixi}
&&\sum_{j=0}^nF_{u_j} (x_i+A_i)_j\nonumber\\
&&
=\sum_{j=0}^nF_{u_j} (\delta_{ij}+A_{ij})\nonumber\\
&&
=\sum_{j=0}^nF_{u_j} \left(\delta_{ij}+A_{iu}u_j
+\sum_{k=0}^nA_{iu_k}u_{kj}\right)\nonumber\\
&&=\sum_{j=0}^nF_{u_j} \left(\delta_{ij}+A_{iu}u_j\right)
+\sum_{k=0}^nA_{iu_k}\sum_{j=0}^nF_{u_j}u_{kj}
\nonumber\\
&&
=0.
\end{eqnarray}
Differentiating Eq. \eqref{Fn} with respect to $x_k$, we have \begin{eqnarray}\label{Fxi}
&&F_k=F_uu_k+\sum_{j=0}^nF_{u_j}u_{jk}=0,
\end{eqnarray}
i.e.,
\begin{eqnarray}\label{Fxi1}
&&\sum_{j=0}^nF_{u_j}u_{jk}=-F_uu_k.
\end{eqnarray}
Substituting Eq. \eqref{Fxi1} into Eq. \eqref{3phixi}, we have
\begin{eqnarray}\label{Eqa}
\sum_{j=0}^nF_{u_j} \left(\delta_{ij}+A_{iu}u_j\right)-\sum_{j=0}^n
A_{iu_j}F_{u}u_{j}
=0.
\end{eqnarray}
Eq. \eqref{Eqa} is a first order linear PDE of $A_i$ with respect to the independent variables $\{u,\ u_i,\ i=0,\ 1,\ \ldots,\ n\}$ and can be solved using the standard characteristic method. The result is merely the one given in Eq. \eqref{3ss} after removing a trivial arbitrary function of $\tau_i$. Theorem 3.3 is proved.$\blacksquare$

In the present section, for arbitrary (1+1)-dimensional first order PDE \eqref{Be}, we have also obtained higher order generalized symmetries by means of the recursion operator(s). However, for higher dimensions, we have not yet find possible recursion operators though there are also infinitely many higher order generalized symmetries for the arbitrary (n+1)-dimensional first order autonomous PDE \eqref{Fn}. In this paper, we will not discuss higher order symmetry of \eqref{Fn} because it is not necessary to find its PBSs as in the (1+1)-dimensional case of the last section.

As mentioned in Section 1, to find PBSs of an (n+1)-dimensional first order PDE, it is enough to use one symmetry with an $n$-dimensional arbitrary function. In this paper, we choose the $n$-dimensional arbitrary function symmetry of the form
\begin{eqnarray}\label{3sn}
\sigma=g(\tau_1,\ \tau_2,\ \ldots,\ \tau_n)u_0,
\end{eqnarray}
where $g$ is an arbitrary function of $\tau_\alpha,\ \alpha=1,\ 2,\ \ldots,\ n$. To find general solutions of \eqref{Fn} via symmetry \eqref{3sn}, we should study the corresponding transformations of $u_i$ because $u_i\ (i=1,\ 2,\ \ldots,\ n)$ enter into the symmetry. Since $u_i\equiv u_{x_i},\ i=0,\ 1,\ 2,\ \ldots,\ n$, we have
\begin{eqnarray}\label{3sui}
\sigma_i&=&(gu_0)_i\nonumber\\
&=&gu_{0i}
+u_0\sum_{\beta=1}^ng_{\tau_\beta}
\left(\frac{u_\beta}{u_0}\right)_{i}
\nonumber\\
&=& gu_{0i}
+u_0\sum_{\beta=1}^ng_{\tau_\beta}\left(\frac{u_{\beta i}}{u_0}
-\frac{u_{\beta}u_{0i}}{u_0^2}\right)
\nonumber\\
&=& gu_{0i}-u_{0i}\sum_{\beta=1}^ng_{\tau_\beta}
\tau_{\beta}
+\sum_{\beta=1}^ng_{\tau_\beta}u_{\beta i}
\nonumber\\
&=&\left(g-\sum_{\beta=1}^ng_{\tau_\beta}
\tau_{\beta}\right)u_{0i}
+\sum_{\beta=1}^ng_{\tau_\beta}u_{\beta i}\equiv Tu_{i0}+\sum_{\beta=1}^n X_\beta u_{i\beta}\nonumber\\
&\equiv&\sum_{j=0}^n X_j u_{ij},\quad X_0=T=g-\sum_{\beta=1}^ng_{\tau_\beta}
\tau_{\beta},\ X_\beta=g_{\tau_\beta},\ \beta=1,\ 2,\ \ldots,\ n.
\end{eqnarray}
It is clear that the expression of symmetry \eqref{3sn} can be rewritten as
\begin{eqnarray}\label{3sna}
\sigma=gu_0=T u_{0}+\sum_{\beta=1}^n X_\beta u_{\beta}=\sum_{j=0}^n X_j u_{j}.
\end{eqnarray}
From Eq. \eqref{3sui} and Eq. \eqref{3sna}, it is known that as in (1+1)-dimensional case, the first prolongation of the symmetry \eqref{3sn} is closed,
\begin{eqnarray}\label{Pr1}
pr^{(1)}\sigma =\left(g-\sum_{\beta=1}^ng_{\tau_\beta}
\tau_{\beta}\right)\partial_t
+\sum_{\beta=1}^ng_{\tau_\beta}\partial_{x_\beta}
+0\cdot\partial_u
+\sum_{\beta=1}^n0\cdot\partial_{u_\beta}.
\end{eqnarray}
Starting from the closed prolongation \eqref{Pr1}, we can prove the following  theorem for the PBSs of \eqref{Fn}.\\
\bf Theorem 3.4. \rm \em
If $u=U(x_i,\ i=0,\ 1,\ \ldots,\ n)$ is a solution of \eqref{Fn}, so is
\begin{equation}\label{run}
u'(x_i,\ i=0,\ 1,\ \ldots,\ n)=U(x'_i(x_j,\ j=0,\ 1,\ \ldots,\ n),\ i=0,\ 1,\ \ldots,\ n)
\end{equation}
with $x'_i\equiv x'_i(x_j,\ j=0,\ 1,\ \ldots,\ n)$ being determined by
\begin{subequations}\label{xn}
\begin{eqnarray}
&&x'_0=x_0-g+\sum_{\beta=1}^n\eta_\beta g_{\eta_\beta},\quad g\equiv g(\eta_1,\ \eta_2,\ \ldots,\ \eta_n),\\
&&x'_\alpha=x_\alpha- g_{\alpha},\ \alpha=1,\ 2,\ \ldots,\ n,
\\
&&\eta_\alpha\equiv \left.\frac{U_\alpha}{U_0}\right|_{x_i\rightarrow x'_i,\ i=0,\ 1,\ \ldots,\ n}.
\end{eqnarray}
\end{subequations}
\it Proof. \rm According to the closed first prolongation \eqref{Pr1}, we can find the finite transformation group via Lie's first principle by solving the initial value problem
\begin{subequations}\label{IVPn}
\begin{eqnarray}
&&\frac{\mbox{\rm d}x_i(\epsilon)}{\mbox{\rm d}\epsilon}
=X_i(\epsilon),\quad x_i(0)=x_i,\quad i=0,\ 1,\ \ldots,\ n,\label{IVPx} \\
&&\frac{\mbox{\rm d}u(\epsilon)}{\mbox{\rm d}\epsilon}
=0,\quad u(0)=U,\label{IVPu}\\
&& \frac{\mbox{\rm d}u_i(\epsilon)}{\mbox{\rm d}\epsilon}
=0, \quad u_i(0)=U_i,\ \quad i=0,\ 1,\ \ldots,\ n.\label{IVPui}
\end{eqnarray}
\end{subequations}
It is clear that from Eq. \eqref{IVPu} and Eq. \eqref{IVPui} we can find
\begin{subequations}\label{uui}
\begin{eqnarray}
&&u(\epsilon)=U(t,\ x_i,\ i=0,\ 1,\ \ldots,\ n),\label{nu} \\
&&u_i(\epsilon)=U_i(t,\ x_i,\ i=0,\ 1,\ \ldots,\ n).\label{nui}
\end{eqnarray}
\end{subequations}
According to the definition of $X_j,\ j=0,\ 1,\ \ldots,\ n$, they are only the functions of $\tau_\beta=u_\beta/u_0$. Thus $X_j(\epsilon)$ are $\epsilon$-independent because of Eq. \eqref{nui}. Owing to this fact, it is trivial to solve Eq. \eqref{IVPx}  resulting in,
\begin{eqnarray}
x_j(\epsilon)=x_j+\epsilon X_j(0).
\end{eqnarray}
Finally, making a notation transformation
$$\{x_j(\epsilon), \ x_j,\ j=0,\ 1,\ \ldots,
\ n\}\longrightarrow \{x_j,\ x'_j,\ j=0,\ 1,\ \ldots,
\ n\}$$
and taking $\epsilon=1$ due to the arbitrariness of $g$, we finish the proof of Theorem 3.4. $\blacksquare$\\
\bf Remark 3.1. \rm As in the proof of Theorem 2.6 in (1+1)-dimensional case, Theorem 3.4 can also be directly proved by calculating $u'_i$ to find the results $u'_i=U_i$. \\
\bf Remark 3.2. \rm For the special selected seed solution $U(x_j,\ j=0,\ 1,\ \ldots,\ n)$, the invariant functions $\eta_j$ may not be functionally independent. We call such kinds of seed solutions as the degenerate seeds. In this case, we have to change the symmetry by including $n$ functionally independent arguments from invariant functions such as $\tau_\alpha,\ \varphi_j,\ \alpha=1,\ 2,\ \ldots,\ n, \ j=0,\ 1,\ 2,\ \ldots,\ n$ and so on. However, in this paper we will not discuss the degenerate seed cases.

\section{Summary and discussions}
In summary, Lie symmetry algebra (group) method is very useful to find solutions of scientific problems. In this paper, the PBSs of arbitrary first order autonomous PDEs can be simply obtained by means of its Lie symmetry algebra. A special type of PBSs of  arbitrary first order autonomous PDEs, which include various important physically important special cases, are given in Theorem 3.4. The only model dependent information is included in their seed solutions. In (1+1)-dimensional case, many other interesting properties such as the recursion operator, invariant operator, higher order symmetries (generators of infinite dimensional Lie symmetry algebra) and invariant functions are also explicitly given.

 It is expected that the symmetry group method proposed here can also be extended to find PBSs of most general nonautonomous first order PDEs which could be solved by means of the so-called complete solution and parameterization method \cite{para}. It is also interesting that the symmetry method can be used to find PBSs and/or complete solutions for higher order nonlinear PDEs. We will report more results of applying the symmetry group method to find PBSs of PDEs in our future researches.\\ \\
\bf Acknowledgement. \rm The author is grateful  to thank Professors R. Conte, C. W. Cao, Y. Q. Li, Y. Chen, Q. P. Liu, X. B. Hu, D. J. Zhang, B. F. Feng, Z. J. Qiao and E. G. Fan for their helpful suggestions and fruitful discussions. The work was sponsored by the National Natural Science Foundations of China (Nos. 11435005, 11471004, 11175092, and 11205092), Shanghai Knowledge Service Platform for Trustworthy Internet of Things (No. ZF1213) and K. C. Wong Magna Fund in Ningbo University.

\small{
}
\end{document}